# Making the Case for Visualization

**Authors:** Robert Hurt (Caltech/IPAC, hurt@ipac.caltech.edu), Ryan Wyatt (California Academy of Sciences, rwyatt@calacademy.org), Mark Subbarao (Adler Planetarium/International Planetarium Society, msubbarao@adlerplanetarium.org), Kimberly Arcand (Chandra X-ray Center, Harvard & Smithsonian Center for Astrophysics, kkowal@cfa.harvard.edu), Jacqueline K. Faherty (American Museum of Natural History, jfaherty@amnh.org), Janice Lee (Caltech/IPAC, janice@ipac.caltech.edu), Brandon Lawton (STScI, lawton@stsci.edu)

## Executive Summary

Visual representation of information is a fundamental tool for advancing our understanding of science. It enables the research community to extract new knowledge from complex datasets, and plays an equally vital role in communicating new results across a spectrum of public audiences.  Visualizations which make research results accessible to the public have been popularized by the press, and are used in formal education, informal learning settings, and all aspects of lifelong learning. In particular, visualizations of astronomical data (hereafter "astrovisualization" or "astroviz") have broadly captured the human imagination, and are in high demand.

Astrovisualization practitioners need a wide variety of specialized skills and expertise spanning multiple disciplines (art, science, technology). As astrophysics research continues to evolve into a more data rich science, astroviz is also evolving from artists conceptions to data-driven visualizations, from two-dimensional images to three-dimensional prints, requiring new skills for development. Currently astroviz practitioners are spread throughout the country. Due to the specialized nature of the field there are seldom enough practitioners at one location to form an effective research group for the exchange of knowledge on best practices and new techniques. Because of the increasing importance of visualization in modern astrophysics, the fact that the astroviz community is small and spread out in disparate locations, and the rapidly evolving nature of this field, we argue for the creation and nurturing of an Astroviz Community of Practice.

We first summarize our recommendations. We then describe the current make-up of astrovisualization practitioners, give an overview of the audiences they serve, and highlight technological considerations.

## Recommendations

1) Established institutions such as NASA, NSF, and Universities should be strongly encouraged to continue to recognize and support visualization expertise which will become increasingly more significant over the next decade.

2) The importance of building a community of practice around astronomy visualization must be prioritized.  This includes increased support for a) visualization-specific workshops and conferences, b) visualization sessions at research and education conferences, and c) professional development for visualization practitioners to develop skills, and engage with others within their community of practice.
3) Open-access policies need to be encouraged so we can develop standards that enable the sharing of visualization assets within and beyond the research and education communities.
4) The visibility and accessibility of visualization tools and associated infrastructure needs to be heightened in the coming decade.  Such tools include stand-alone software used in research and outreach, code bases and libraries that enable custom development, and adaptations of general commercial and off the shelf software.
5) Funding opportunities for visualization software tools that benefit wide cross-sections of the community and support the continued improvement of existing infrastructures should be prioritized.

# Key Points

## Astrovisualization practitioners: at the intersection of art, science, technology, communication, and education

In astrophysics, the communities of practice that participate in visualizing data include the following:

1. **Science community:** For the science community, there is an ecosystem of software development around tools that the community can use to find archival data and visualize astrophysics data. Examples include **glue**, **Aladin**, and **WorldWide Telescope**, among others. This ecosystem includes practitioners that create the tools and use the tools for scientific discovery and dissemination.

2. **Communications teams from astrophysics missions:** Communications teams from astrophysics missions create a wealth of astrophysics visualizations for non-experts. The goals of this group include making complicated physical phenomena accessible and promoting the capabilities and value of astrophysics missions. These missions tend to have astronomers working with the visualization teams to make sure that the science content is accurate. Many astrophysics missions, such as NASA's Great Observatories, have their own visualization experts that create visualizations to help increase awareness and understanding of difficult science concepts and the mission's role in furthering our knowledge of those concepts (e.g., see DePasquale et al 2015; Hurt & Christensen 2007; Hurt 2010; Rector et al 2017, 2007).

3. **Education teams from astrophysics missions:** Education teams create and provide astrophysics visualization content for informal learning venues, educators, and developers, as well as lifelong learners. Resulting visualizations might appear in informal learning venues such as planetariums, science and tech museums (theater/dome/exhibit), libraries, and other places of informal and lifelong learning, including online. These resources are shared with informal educators via professional learning experiences and communities of practice, as well as through our websites and social media (including YouTube).

4. **Planetarium and museum communities:** The museum and planetarium communities serve as both content consumers and creators. The specialized nature of the planetarium often requires specific formatting for visualization content. Several planetariums (e.g., Hayden Planetarium in New York, Adler Planetarium in Chicago, and Morrison Planetarium in San Francisco) create visualization-based planetarium content that is then distributed globally. But virtually all modern planetariums offer real-time visualization platforms that can leverage research data sets; indeed, planetariums could offer professional astronomers a visualization environment that can drive astronomical discovery (see the white paper from Faherty et al., "IDEAS: Immersive Dome Experiences for Accelerating Science").

5. **Amateur and enthusiast communities:** There are multiple amateur and enthusiast communities that have sprung up to use astrophysics visual resources. These include night sky enthusiasts, astrophotographers, amateur image processors, citizen science projects, and educators that want to use astrophysics examples to teach examples of physical phenomena to leverage the excitement that the youth has around space. These amateur/enthusiast communities rely upon the visuals, tools, and data from astrophysics missions, astrophysics communications teams, and educational programs. By sharing these visualizations and tools more broadly, the amateur and enthusiast communities are increasing the reach of astrophysics science.

These seemingly disparate communities find common ground in the challenges they face, technical and otherwise, in producing visualizations to accomplish their various and distinct goals. Since 2005, the "AstroViz" community has gathered in places ranging from Chicago to Seattle to San Francisco to Pasadena, in order to discuss content and share ideas. This impromptu and self-organized gathering occurred annually through 2010 but experienced a significant gap until June 2018, when a workshop sponsored by NASA's Universe of Learning[1] reconvened the community. Sponsoring such recurring gatherings—as well as sessions at established research and education conferences—would be a critical step in ensuring the longevity and robustness of the community of practice.

---

[1] We also highlight the efforts of the NASA Science Activation Visualization Working Group, the authors are sharing these ideas with and among that group as well.

## Astrovisualization audiences

Even more so than when working with mission planning or the astrophysics science community, the education/outreach community presents a diverse community (general public, educators at informal learning venues, lifelong learners, pre-service teachers, etc.) with diverse needs and learning objectives. Understanding the requirements of the target audiences is essential to developing effective visualizations that meet communications or learning goals, and recognizing the necessity of advancing inclusive techniques from data sonification to 3D printing (see for example Diaz-Merced et al 2012; Diaz-Merced 2013, 2014; Christian et al 2015; Madura 2017; Madura et al; Steffen et al 2011, 2014; Grice et al 2015).

## Visualization for Research

As astronomical data volume continues expanding dramatically, tools for visualizing these datasets will become an increasingly essential asset to support both basic research and professional communication. Exploration of vast datasets, whether they cover large areas, extend over long time sequences, or encompass multi-dimensional phase spaces of derived properties, will continue to be a critical aspect of analysis and discovery. Likewise, techniques to visually present results in publications and talks will play a critical role in the effective communication of science within the community.

Recognition of the importance of visualization techniques and tools, and corresponding support for research and development, will play a vital role in making the most of valuable research data. A research climate supportive of the development and sharing of robust visualization tools and code libraries can enhance productivity across the field. And finally, building effective communities of practice will permit sharing assets and expertise across disciplines that can benefit the entire astronomy community.

## Visualization for Education & Public Outreach

Engaging public interest in astronomy can achieve many positive outcomes, ranging from improving core STEM education and enhancing the workforce to increasing critical thinking skills, public awareness, and support for basic astronomy research. The efficacy of such efforts over the past two decades can be tied in no small part to the power of astronomical visualizations that reach broad audiences in print, on the web, in classrooms, and in informal learning environments such as museums and planetariums.

In the news media, astronomy has fared particularly well among sciences in receiving significant coverage in a deeply-oversubscribed news environment. This is often achieved with something as simple as a carefully crafted astronomical data image (Figure 1). Likewise, science-based illustrations have proven vital to news coverage of technical astronomy results, serving both to stimulate awe and curiosity in and to foster an understanding of the underlying science.

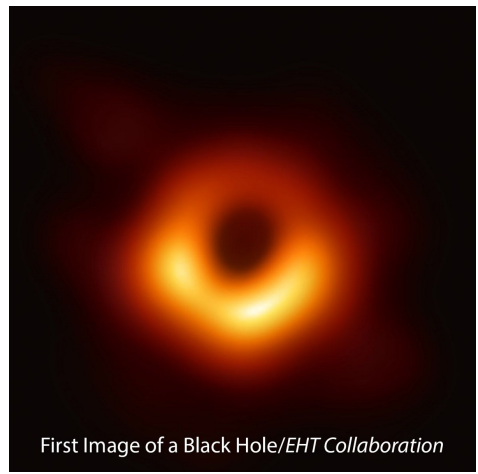

First Image of a Black Hole/*EHT Collaboration*

*Figure 1.* Examples of recent news stories that have successfully leveraged visualizations, reaching audiences of millions.

Museums and planetariums, key facilities for informal education, are likewise deeply dependant upon effective visualization to interest and educate patrons. Their assets can span the full range of visualization types: static imagery, videos and animations, immersive environments, real-time and interactive data display, and increasingly augmented- and virtual-reality experiences. Museums and planetariums reach a large worldwide audience. There are approximately 3,000 science centers in the world serving an audience of roughly 310 million people each year (Mechelen Declaration 2014). A rough estimate gives 4,000 planetariums in the world, with an annual attendance of 150 million (Peterson 2018).

## Creators, Mediators, and Consumers

Data in an archive has potential value, but data that has been visualized—for a specific audience—has an additional activated value. Knowing the targeted audience(s), their settings, and their use cases, therefore, is an important aspect of any scientific visualization creation and dissemination pipeline (Arcand et al. 2013; Smith et al. 2015). Often scientific data is complex, multidimensional or multimodal which can be impediments to the user for making meaning (Arcand et al. 2013, 2017; Frankel 2004; Smith et al. 2011). Therefore, visualizations designed to align with the users' needs is critical across levels of impact and engagement.

In this section, we enumerate three groups that span stakeholders in visualization: creators, mediators, and end users. Please note that many users, however, will not fit cleanly in a single category but stretch across the spectrum.

Creators of visualizations include subject matter experts (e.g., scientists, programmers, technologists, data visualizers, artists), citizen scientists, amateur astronomers, educators and outreach professionals (e.g., EPO teams, planetarium producers), and students. Mediators of visualizations include formal educators (e.g., in K–16 environments), informal educators (e.g., at museums, science centers, libraries, afterschool programs), amateur astronomers, public media, and social media influencers. End users and consumers of visualizations include public media, social media users, science-attentive and -interested publics, educators, students, and non-experts.

Community efforts have primarily focused on astrovisualization creators, but future efforts should focus on the needs of creators, mediators, and consumers.

## Technological developments

Scientific visualization techniques are evolving at an accelerating rate. Rapid developments in computer graphics, driven primarily for movies and video games, have enabled scientific graphics to achieve a level of detail never before seen. Astronomy, being the most visual of the sciences, has benefited more than any other from computer graphics advances. Visualization techniques and tools will need to continue to advance in the era of big data as we seek to analyze datasets covering larger areas at higher resolutions at higher temporal cadences.

In addition, the current transition from traditional opto-mechanical planetariums to fulldome digital theaters has increased the demand, supply, and quality of astronomical visualizations and moved them from flat screen to immersive environments (Figure 2). And now the growing market of virtual and augmented reality devices has the potential to grow the opportunities for immersive experiences to include personal interactivity with potential applications ranging from outreach to research.

While planetariums have historically been the singular focus for immersive experiences at an audience level, advances in virtual and augmented reality hardware offer intriguing opportunities for building on that legacy and offering new modalities for both outreach and data analysis. Immersive assets can take many forms, from simple hemispheric (fulldome) or 360° video, to software that allows projection and manipulation of 3D datasets and assets in real time. There are potential synergies to be explored across multiple industries; for instance, the same software platforms developed for video games have proven useful for astronomical assets. Just as the VR/AR gaming industry is very much in its nascent stages, the true potential for immersive applications for astronomical data analysis, communications and outreach has yet to be explored fully (see Ferrand, English, & Irani, 2016; Arcand et al 2018).

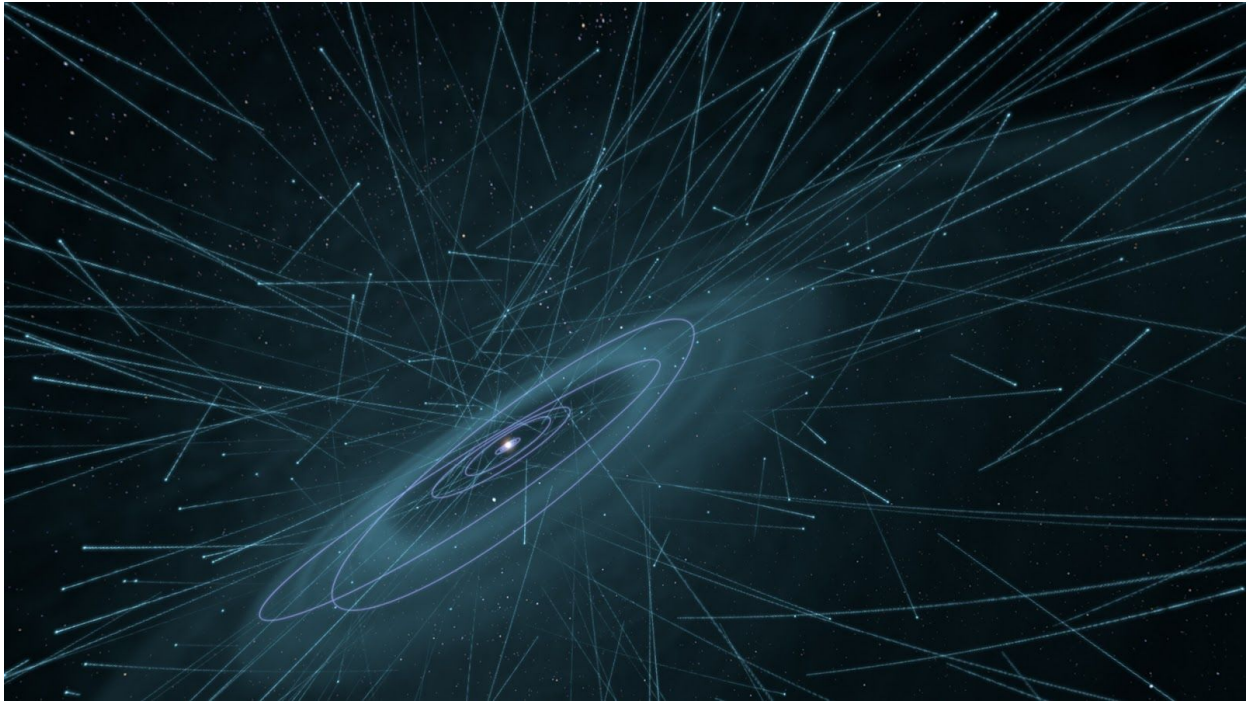

*Figure 2.* A still image from a planetarium show produced by the California Academy of Sciences, depicting the dynamical evolution of Kuiper Belt Objects in the early history of the Solar System based on simulation data (Batygin & Brown 2011). This image has been reformatted from the fulldome "fisheye" view projected on a planetarium dome.

The infrastructure to share visual assets across astrophysics would greatly benefit the greater astronomy visualization community. This was a need identified at the 2018 AstroViz workshop.

It is critical to understand the suite of existing resources available for visualizing data. For astrophysics, direct connections to the science, scientists, and science data and tools allows us to leverage existing infrastructure and data tools and augment them to be used by our audiences. By doing so, we can meet the needs of our audiences and our learning goals with modest investment.

The EPO archives of Hubble, Spitzer, Chandra, and NRAO have benefited from adopting the astronomy visualization metadata (AVM) standards for astronomical imagery. It is now easier for our audiences—including planetarians and educators—to search for images and learn science content while doing so. A planetarium content discovery standard (data2dome) provides a mechanism for content providers to deliver content directly to planetariums. On the same day as the press release of a major discovery, planetarium operators will find planetarium content related to that discovery on their console. This includes not only the visual assets to display on the dome, but also the contextual information necessary to interpret them effectively.

While there is a growing community of amateurs and enthusiasts creating visualizations, we recognize the unique ability for astrophysics missions and science education programs to provide timely and accurate visualizations to our audiences at the same time that news articles are released by communications or other scientific teams. The direct connection to the science, scientists, and data allows for the development of experiences for learners that explore cutting edge new science endeavors.

# Strategic Plan

Astrovisualization plays a key role in supporting both research and education. In the next ten years, the astronomy community needs to enhance its efforts to enhance visualization opportunities for all audiences.

We must continue to develop tools and infrastructure to support visualization for all audiences. In particular, the demands of big data (from Gaia, LSST, WFIRST, etc.) will force the development of new tools and techniques for visualization. These tools and techniques must support browsing multi-dimensional catalog data efficiently, analysis of large-scale high dynamic range imagery and spectroscopic surveys, and time-domain handling of all of the above. Standard imaging software is ill-equipped to handle the community's growing needs, and new technologies like augmented/virtual reality offer significant potential for research and communication visualizations.

We must recognize the significance of sharing visualization assets within and beyond the research community. This requires the development of standards to facilitate sharing and tools that support interoperability between research and education. We want to avoid reinventing wheels, so encouraging interoperability of existing, specialized toolkits will allow the community to leverage successful work that has already been achieved. Furthermore, we must advocate for adoption of open-access policies as much as feasible for visualization (much in the same way that access is supported for data) and continue to support the existing archives as critical infrastructure that supports visualization efforts.

Finally, we must recognize the importance of supporting the astroviz community of practice. This can be accomplished through visualization-specific workshops and conferences, sessions at research and education conferences, and professional development opportunities for professional astronomers and visualization specialists. This can stimulate innovation and creative problem-solving that can only be accomplished within dynamic communities.

# Notes & References